\documentclass[aps,pre,twocolumn,showpacs,showkey,square,numbers,amssymb,amsmath,nobibnotes,mathtools,longbibliography]{revtex4-1}
\usepackage{bm}
\usepackage{times,float}
\usepackage{graphicx}
\usepackage[usenames,dvipsnames,svgnames]{xcolor}
\usepackage{hyperref}
\hypersetup{colorlinks=true, linkcolor=NavyBlue, citecolor=PineGreen,urlcolor=cyan}

\usepackage{color}
\definecolor{blue}{rgb}{0,0,1}

\begin{document}
\title{Statistical properties of speckle patterns for a random number of scatterers and non-uniform phase distributions}
\author{Fernando L. Metz, Cristian Bonatto and Sandra D. Prado}
\address{Physics Institute, Federal University of Rio Grande do Sul, 91501-970 Porto Alegre, Brazil}

\begin{abstract}
The statistical properties of speckle patterns have important applications in optics, oceanography, and
transport phenomena in disordered systems. Here we obtain closed-form analytic results for the amplitude distribution of speckle patterns
formed by a random number of partial waves characterized by an arbitrary phase distribution, generalizing classical results of the random
walk theory of speckle patterns. We show that the functional form of the amplitude distribution is solely determined
by the distribution of the number of scatterers, while
the phase distribution only influences the scale parameters.
In the case of a non-random number of scatterers, we find an analytic expression for the amplitude distribution that extends
the Rayleigh law to non-uniform random phases.
For a negative binomial distribution of the number of scatterers, our results reveal that large fluctuations of the wave amplitudes become more pronounced
in the case of biased random phases. We present numerical results that fully
support our analytic findings.
\end{abstract}
\maketitle

\section{Introduction}

Waves propagating in random media undergo multiple scattering from inhomogeneities \cite{IshimaruBook}. The scattered waves interfere
with each other and give rise to an irregular intensity pattern that echoes the random structure of the sample.
In optics, the most straightforward realization of this basic phenomenology occurs when a laser beam is
reflected by a rough surface, leading to the formation of a granular intensity pattern known
as a speckle pattern \cite{Goodman,GoodmanBook,Carminati2021}.
Studies of various aspects linked to speckle formation have been carried out since the laser
  invention and remain an active area of research \cite{Dogariu2015,Carminati2021}. Prominent statistical properties include not only
  the intensity distribution but also other speckle distributions, such as those related to light polarization \cite{Cohen1991} and phase singularities \cite{Freund1994}.
  Another important topic that has been experiencing a rapid development, both from a theoretical and an experimental point of view, is the study of 
  three-dimensional speckle patterns \cite{2019Skipetrov,2021Leonetti,2022Ott}.
The statistical properties of speckle patterns find important applications in hydrodynamics \cite{Dudley2019}, medical science \cite{2009Draijer}, imaging
techniques \cite{2006Gatti,2016Shin,2020Ye}, material science \cite{Valent2018,Castilho2023}, multimode
fiber systems \cite{SeniorBook}, optical radar
performance \cite{OscheBook}, and transport phenomena in disordered systems \cite{Mirlin2000,Wang2011}.

The random walk model is the simplest theoretical approach
to study speckle patterns \cite{Goodman,GoodmanBook}. In this framework, the resultant electromagnetic field
observed at a specific point
is the superposition of a large number of partials waves, each one arising from an individual scatterer within the sample.
Each contribution to the speckle field carries a random phase, whose statistical properties reflect the spatial
arrangement of the scatterers. The problem is formally equivalent to a random walk in two dimensions \cite{Jakeman1987,GoodmanBook}.
When the phases are independent random variables, the central limit theorem
holds and the resultant electromagnetic field follows a Gaussian distribution. In the strong scattering regime, where the phases are uniformly
distributed, the amplitude of the resultant field follows the well-known Rayleigh distribution \cite{Goodman,GoodmanBook}.
In the weak scattering regime, in which the phases exhibit weak fluctuations around a constant value, the amplitude obeys
the Rice distribution \cite{GoodmanBook}.

The breakdown of the central limit theorem in the random walk model leads to deviations from both
the Rayleigh and the Rice distribution.
Indeed, non-Rayleigh statistics emerges in the strong scattering regime as long as the phases are correlated random
variables \cite{Bromberg2014,Bender2018,Bender2019,Bonatto2020,daSilva2020} or the number of scatterers is finite \cite{Pusey1974,Popov1993,Watts1996}.
More fundamental approaches, based on the solutions of the classical wave equation \cite{Shapiro1986,Mirlin2000}, have shown
that the exponential tail of the Rayleigh distribution
fails to reproduce the statistics of large amplitudes \cite{Dashen1979,Mirlin1998,Mirlin2000}.
In the context of random walk models, Jakeman and Pusey \cite{Jakeman1978}
put forward a phenomenological model for the strong scattering regime, in which the number of scatterers is itself a random variable.
When the variance of the number of scatterers is large enough, the amplitude of the resultant speckle field follows the
so-called $K$-distribution, which has been recognized as a very good model 
for scattering experiments involving turbulent media \cite{Phillips1981,Jakeman1988}.
From a more theoretical perspective, the $K$-distribution can be seen as a consequence of the breakdown of the central limit
theorem, which occurs due to the fluctuating number of scatterers \cite{Jakeman1978,Jakeman1980,Jakeman1987,Jakeman1988}.

Theoretical approaches for speckle patterns have traditionally focused on the limiting cases of strong and weak scattering regimes. The
Rayleigh distribution characterizes amplitude fluctuations resulting from a highly inhomogeneous medium, while the Rice distribution arises from a quasi-homogeneous medium.
In these two limits, the phases of the partial waves are uniformly distributed. On the other hand, much
less attention has been devoted to the scattering of waves with nonuniform phase distributions, i.e., physical
situations in which the values of the phases have different statistical weights. These cases are of physical relevance
since nonuniform phase distributions play an important role in a number of
practical situations, such as the scattering
from a finite disordered sample \cite{Kazmierczak1989}, the speckle patterns formed in the
near-field region \cite{Apostol2003,Apostol2005,Apostol2006}, and the amplitude fluctuations
arising from non-isotropic arrangements of scatterers \cite{Naraghi2016}.

Concerning nonuniform random phases, there have been a few attempts  \cite{Barakat86,Jakeman1987,Jakeman1988} to generalize
the random walk model of speckle patterns to include this feature. 
Barakat \cite{Barakat86} has derived the amplitude distribution in the particular case
of a Von Mises phase distribution, while Jakeman and Tough \cite{Jakeman1987,Jakeman1988} have considered
a weak perturbation around the uniform phase distribution. Despite the fact that many results from the classical speckle
theory have been developed long time ago, the interest in the statistical properties of both linear and nonlinear wave propagation has been renewed with
the growing interest in extreme wave generation \cite{Arecchi2011,Mathis2015,Bonatto2020,Safari2017,Tlidi2022}.
To the best of our knowledge, there is no
systematic approach to calculate the amplitude distribution of speckle patterns for a generic distribution of phases. 
The purpose of our paper is to fill this gap.

Here we introduce an exactly solvable random walk model 
for speckle patterns generated
by a random number of scatterers with an arbitrary phase distribution. By distinguishing between
biased and unbiased random phases, we derive general analytic expressions
for the amplitude distribution of the speckle field, from which one can find
several closed-form analytic results that cover a rich variety of specific situations, generalizing
classical results \cite{Goodman,Jakeman1978} within the theory of speckle patterns.
In particular, we find
a novel analytic result for the amplitude distribution in the simple case of a non-random number of scatterers, which extends
the Rayleigh law to arbitrary phase distributions.
We show that the distribution of the number
of scatterers fully determines the functional form of the amplitude distribution, while the information about the phase distribution is encoded 
in the scale parameters characterizing the joint distribution of the speckle field.
For a negative binomial distribution of the number of scatterers, we show that amplitude fluctuations become more
pronounced in the case of biased random phases, due to a power-law tail in the amplitude distribution.
The exactness of our theoretical findings are fully confirmed by numerical simulations.

The paper is organized as follows. In the next section we introduce the random walk model for
the scattering by an inhomogeneous surface with a random number of scatterers. Section \ref{Intens} presents
the main analytic results for the amplitude distribution in the cases of biased and unbiased random phases. In
section \ref{Final}, we summarize our results and discuss some perspectives of future work. The paper contains
two appendices. Appendix \ref{calc} explains all details involved in the derivation of the
main analytic findings of section \ref{Intens}, while in appendix \ref{calc1} we demonstrate
an useful identity to obtain closed-form expressions for the amplitude distribution when
the number of scatterers follows a negative binomial distribution.

%%%%%%%%%%%%%%%%%%%%%%%%%%%%%%%%%%%%%%%%%%%%%%%%%%%%%%%%%%%%%%%%%%%%%%%

\section{The random walk model for speckle patterns}

In this work we focus on the random walk model of speckle patterns. One of the original
 motivations for introducing this phenomenological
 model is the study of the interference pattern formed at a large distance from an irregular surface \cite{Goodman,Jakeman1988,GoodmanBook}.
For concreteness, we consider a one-dimensional regular lattice of length $L$ with $N$ sites
or positions labeled by $j=0,\dots,N-1$, in which the distance between two adjacent sites is given by $\frac{L}{N-1}$.
Each site is occupied by a pointlike source that emits coherent light with wavelength $\lambda$. We will study the distribution
of the amplitude that characterizes the resulting interference pattern emerging at a large distance from the sample (Fraunhofer diffraction).
Since pointlike scatterers emit spherically symmetric partial waves, we assume that the amplitudes of all individual waves are equal to a constant $E_0$.
Let $\phi_j$ ($j=0,\dots,N-1$) be the phase of the partial wave arising from  the $j$ scatterer.
The electromagnetic field on a screen or observation plane located at a large distance from the sample is given by
\begin{equation}
  E(\beta) = E_0 \sum_{j=0}^{N-1} e^{i j \frac{\beta}{N-1} + i \phi_j },
  \label{hopi}
\end{equation}
where 
\begin{equation}
\beta = 2 \pi \frac{L}{\lambda} \sin{(\theta)}
\end{equation}
is given in terms of the angular
position $\theta \in (-\frac{\pi}{2},\frac{\pi}{2})$ on the screen.

Equation (\ref{hopi}) holds when all partial waves arrive on the screen. There are different reasons
to drop this assumption and consider a more realistic scenario in which light is emitted
only by a fraction of scatterers randomly placed over the sample. This class of models
is particularly relevant when the scatterers exhibit dynamic behavior, moving
in and out of the illuminated region due to dynamical processes \cite{Jakeman1980}.
A minimal microscopic model of this situation is represented by the following expression for the electric field
\begin{equation}
  E(\beta) = E_0 \sum_{j=0}^{N-1} x_j e^{i j \frac{\beta}{N-1} + i \phi_j },
  \label{hopi1}
\end{equation}
where the binary random variable $x_j \in \{ 0, 1 \}$ tells whether the individual wave labeled by $j$ reaches the detector. If $x_j=1$, then
the $j$-wavefront reaches the screen, whereas $x_j=0$ otherwise. Therefore, the occupation random variables $x_{0},\dots,x_{N-1}$
determine the (random) positions of the scatterers along the one-dimensional lattice. 
The total number $k=0,\dots,N-1$ of wavefronts that reach the screen is now given by
\begin{equation}
k = \sum_{j=0}^{N-1} x_j.
\end{equation}
Note that the speckle field in Eq. (\ref{hopi1}) is a sum of $k$ terms. The phases $\phi_0,\dots,\phi_{N-1}$ are independent
random variables that follow an arbitrary distribution $\Omega(\phi)$.

In order to complete the definition of the model, we need to specify the distribution
of $x_0,\dots,x_{N-1}$. The discrete quantity $k=0,\dots,N-1$ is itself a random variable that models the microscopic
fluctuations of the number of scatterers among different samples.
Given a certain value of $k$, we choose the conditional joint distribution $P(\boldsymbol{x} |k )$ of  $\boldsymbol{x} = (x_0,\dots,x_{N-1})$ as follows
\begin{align}
  P(\boldsymbol{x}|k ) &= \frac{1}{\mathcal{N}_k} \delta_{k, \sum_{j=0}^{N-1} x_j  }  \nonumber  \\
  &\times \prod_{j=0}^{N-1} \left[ \frac{k}{N} \delta_{x_j,1} + \left(1 - \frac{k}{N}  \right) \delta_{x_j,0} \right],
  \label{rigt}
\end{align}  
where $\mathcal{N}_k$ is the normalization factor for finite $N$. In the limit $N \rightarrow \infty$, this factor converges to $\mathcal{N}_k^{(\infty)} = \frac{e^{-k} k^k}{k!}$.
According to Eq. (\ref{rigt}), the variables $x_0,\dots,x_{N-1}$ are independently
drawn from a bimodal distribution, where $x_j =1$ with
probability $k/N$ and $x_j =0$ with probability $1 - k/N$. The term containing the Kronecker $\delta$ ensures
that the constraint $k=\sum_{j=0}^{N-1} x_j$ is fulfilled for each realization of the model.
The object $P(\boldsymbol{x} |k )$ is the conditional probability
for a fixed $k$, while the joint distribution $p(\boldsymbol{x})$ is obtained from
\begin{equation}
  p(\boldsymbol{x}) = \sum_{k=0}^N p_k P(\boldsymbol{x} |k ),
  \label{rigt1}
\end{equation} 
where $p_k$ is the discrete probability of $k$. Equations (\ref{rigt}) and (\ref{rigt1}) completely
define the distribution of $x_0,\dots,x_{N-1}$.

Our primary aim is to compute the probability distribution $\mathcal{P}_{\beta}(A)$ of the amplitude
\begin{equation}
  A(\beta) = | E(\beta) |,
  \label{ghop}
\end{equation}  
generalizing classical results \cite{Goodman,Jakeman1978} of the theory of speckle patterns
to arbitrary phase distributions $\Omega(\phi)$. Once the analytic form of $\mathcal{P}_{\beta}(A)$ is known, the distribution $\mathcal{F}_{\beta} (I)$
of the intensity $I = A^2$ is determined by the relation $\mathcal{F}_{\beta} (I) = (4 I)^{-1/2}  \mathcal{P}_{\beta}(\sqrt{I})$, which is obtained
by a simple change of variables.
Since $N$ is the number of available positions of the scatterers along the sample, the average density of scatterers is given by
\begin{equation}
D = \frac{a}{N},
\end{equation}  
where $a$ is the average number of scatterers
\begin{equation}
a = \sum_{k=0}^N p_k k.
\end{equation}  
In the next section, we present analytic expressions for $\mathcal{P}_{\beta}(A)$ in the regime where both $N$ and $a$ are infinitely large, but the density
$D$ goes to zero.
This low density regime is achieved by setting
$a \propto N^{\delta}$ ($\delta < 1$) and then taking the limit $N \rightarrow \infty$. Put differently, our analytic findings are valid  in the regime
where the average number of scatterers is very large, but much smaller than the total number of available space in the sample.

%%%%%%%%%%%%%%%%%%%%%%%%%%%%%%%%%%%%%%%%%%%%
%%%%%%%%%%%%%%%%%%%%%%%%%%%%%%%%%%%%%%%%%%%%%

\section{The distribution of amplitudes}
\label{Intens}

In this section we present the main analytic results for $\mathcal{P}_{\beta}(A)$. Let $\mathcal{W}_{\beta}(E_R,E_I)$
  be the joint probability distribution of the complex field $E(\beta)= E_R(\beta) + i E_I(\beta)$ for fixed $\beta$.
In order to obtain a finite limit of $\mathcal{W}_{\beta}(E_R,E_I)$ as $a \rightarrow \infty$, we need
to rescale the amplitude $E_0$ with respect to $a$. The rescaling factor depends
on the choice of the distribution $\Omega (\phi)$ of phases $\phi_0,\dots,\phi_{N-1}$.
Thus, although there is no need to fully specify the form of $\Omega(\phi)$ to derive the expressions for $\mathcal{P}_{\beta}(A)$, we do have to distinguish
between two different families of distributions, since $E_0$ has to be rescaled differently in each case.
Let $\langle f(\phi) \rangle_{\phi}$ be the average of a function $f(\phi)$ of the random phase $\phi$, 
\begin{equation}
  \langle f(\phi) \rangle_{\phi} = \int_{0}^{2 \pi} d \phi \Omega (\phi) f(\phi).
  \label{skcm}
\end{equation}  
For unbiased random phases, $\Omega(\phi)$ is such that $\langle e^{i \phi} \rangle_{\phi} =0$ and we rescale
the amplitude of the speckle field as $E_0 \rightarrow E_0/\sqrt{a}$. An important example of an unbiased distribution $\Omega(\phi)$ is the
uniform distribution in the interval $[0,2 \pi)$, which characterizes the strong scattering regime.
For biased random phases, $\Omega(\phi)$ is
such that $\langle e^{i \phi} \rangle_{\phi} \neq 0$ and the amplitude must be rescaled as $E_0 \rightarrow E_0/a$.
The most straightforward example of biased phases arises when all phases are set to a constant value. 
 
Apart from the aforementioned constraints on $\Omega(\phi)$, our main analytic results hold for arbitrary distributions $p_k$
and $\Omega(\phi)$. In other words, both the distribution of the number of scatterers and
the phase distribution are inputs of the analytic expressions. Nevertheless, for the purpose of generating numerical results, we do have to
specify $p_k$ and $\Omega(\phi)$. As a simple example of a non-uniform $\Omega(\phi)$, we will consider a bimodal distribution
\begin{equation}
  \Omega(\phi) = q \delta(\phi-\phi_0) + (1-q) \delta(\phi-\phi_0-\pi),
  \label{ggsj}
\end{equation} 
where $\phi_0 \in [0,2 \pi)$ and $q \in [0,1]$. According to Eq. (\ref{ggsj}), we randomly select phases $\phi_0$ and $\phi_0 + \pi$ with
  probabilities $q$ and $1-q$, respectively. When $q=1/2$, Eq. (\ref{ggsj}) produces an unbiased distribution, while
  for $q \neq 1/2$, the distribution $\Omega(\phi)$ becomes biased. 

As we will see below, when $a$ approaches infinity, the statistics of the number of scatterers
is encoded in the function
\begin{equation}
  \nu(g) = \lim_{a \rightarrow \infty} \sum_{k=0}^N p_k \delta \left(g - \frac{k}{a} \right),
  \label{hgds8}
\end{equation}  
which  provides the distribution of the rescaled number $k/a$ of scatterers when $a \rightarrow \infty$. The shape
of $\nu(g)$ is dictated by the discrete distribution $p_k$. When discussing explicit results for the amplitude 
distribution, it is convenient to distinguish between two regimes that characterize the fluctuations of the number $k$ of scatterers. Let $\Delta^2_\nu$ be the
variance of $\nu(g)$. In the regime of {\it weak} fluctuations of $k$, the discrete distribution $p_k$ is such
that $\nu(g) = \delta(g-1)$ and, therefore, $\Delta_\nu=0$. A typical example where $p_k$ results in vanishing fluctuations in the number of scatterers
is given by the Poisson distribution $p_k = \frac{a^k e^{- a}}{k!}$. In the regime of {\it strong} fluctuations of $k$, $p_k$ is such
that $\nu(g)$ has a finite variance ($\Delta_\nu > 0$).

We model the regime of strong fluctuations by following \cite{Jakeman1978} and considering a negative binomial distribution of
the number of scatterers 
\begin{equation}
  p_k = \frac{\Gamma(\mu + k) }{\Gamma(\mu)} \frac{1}{k!} \left( \frac{a}{\mu}  \right)^k  \frac{1}{\left(1 + a/\mu   \right)^{\mu + k} },
  \label{fsdp}
\end{equation}  
where the parameter $\mu > 0$ controls the variance $\sigma_k^2$ of $p_k$ according to
\begin{equation}
\sigma_k^2 = a + \frac{a^2}{\mu}.
\end{equation}  
As $\mu \rightarrow \infty$, Eq. (\ref{fsdp}) converges to a Poisson distribution with mean $a$ and we recover the regime of weak fluctuations. For $\mu =1$, Eq. (\ref{fsdp}) gives
rise to the geometric (or exponential) distribution
\begin{equation}
p_k = \frac{1}{a+1} \left( \frac{a}{a+1}  \right)^k.
\end{equation}
From a practical perspective, the negative binomial distribution
  can be interesting to study the scattering by samples in which spatial correlations in the positions of the scatterers
  induce the formation of clusters of fluctuating size. If the size $L$ of the scattering region is comparable to the
correlation length, the number of scatterers inside the region defined by $L$ should display large fluctuations. This argument has motivated
the use of the negative binomial distribution in random walk models that fit empirical data obtained from scatering experiments with a variety of turbulent systems \cite{Jakeman1978,Jakeman1988}.

Inserting Eq. (\ref{fsdp}) into Eq. (\ref{hgds8}), one can show that the corresponding $\nu(g)$ is the gamma distribution
\begin{equation}
  \nu(g) = \frac{\mu^\mu}{\Gamma(\mu)} g^{\mu -1} e^{-\mu g},
  \label{gfdsw}
\end{equation}  
whose variance is given by
\begin{equation}
  \Delta_{\nu}^{2} = \frac{1}{\mu}.
  \label{hhas}
\end{equation}  
The above equation further clarifies why this model is interesting. The gamma distribution, Eq. (\ref{gfdsw}), interpolates between the regimes
of weak fluctuations ($\mu \rightarrow \infty$) and strong fluctuations ($\mu \rightarrow 0$) by only changing a single parameter $\mu$. As expected, Eq. (\ref{gfdsw}) reduces to the exponential
distribution $\nu(g) = e^{-g}$ for $\mu=1$. 

In summary, the distribution $\nu(g)$, analogous
to the continuous version of $p_k$, along with $\Omega(\phi)$, determines the shape of the amplitude distribution $\mathcal{P}_{\beta}(A)$.
In the next subsections, we discuss explicit results where $\Omega(\phi)$ and $\nu(g)$ follow, respectively, Eqs. (\ref{ggsj}) and (\ref{gfdsw}).

%%%%%%%%%%%%%%%%%%%%%%%%%%%%%%%%%%%%%%%%%%%%%%%%%%%%%%

\subsection{Biased phases}

First, we present the analytic results for distributions of phases that fulfill $\langle e^{i \phi} \rangle_{\phi}\neq 0$.
In this case, the joint distribution $\mathcal{W}_{\beta}(E_R,E_I)$ of the complex field $E(\beta)=E_R(\beta) + i E_I(\beta)$, Eq. (\ref{hopi1}), is given by
\begin{eqnarray}
  \mathcal{W}_{\beta}(E_R,E_I) &=& \int_{0}^{\infty} d g \, \nu(g) \delta\left[ E_R - g \, {\rm Re} E_{*}(\beta)  \right] \nonumber \\
  &\times&
  \delta\left[ E_I -  g \, {\rm Im}  E_{*}(\beta)    \right],
  \label{kkap}
\end{eqnarray}
where the mean value $E_{*}(\beta)$ of the speckle field is a function of the position $\beta$ on the screen, namely
\begin{align}
  E_{*}(\beta) &= \frac{E_0}{\beta}  \left\langle \sin(\beta + \phi) - \sin \phi  \right\rangle_{\phi} \nonumber \\
  &+ i \frac{E_0}{\beta}  \left\langle  \cos \phi - \cos(\beta + \phi)    \right\rangle_{\phi}.
  \label{jjaa1}
\end{align}
Equation (\ref{kkap}) is valid when both $N$ and $a$ become infinitely large, but the average density $D = a/N$ of the number
of scatterers goes to zero.

The distribution $\mathcal{P}_{\beta} (A)$ of the amplitude readily follows from Eqs. (\ref{ghop}) and (\ref{kkap}),
\begin{equation}
  \mathcal{P}_{\beta} (A) = \frac{1}{|E_{*}(\beta)| } \nu\left( \frac{A}{|E_{*}(\beta)| }  \right).
  \label{gdew1}
\end{equation}
The above result holds for any distribution $\Omega(\phi)$ of phases, provided $\langle e^{i \phi} \rangle_\phi \neq 0$, and for
any distribution $\nu(g)$ of the rescaled number of scatterers. The distribution $\nu(g)$ controls the functional form
of $\mathcal{P}_{\beta} (A)$, while $\Omega(\phi)$ appears in the scale parameter $|E_{*}(\beta)|$. We explain how to derive
Eqs. (\ref{jjaa1}) and (\ref{gdew1}) in appendix \ref{calc}.

The moments of the amplitude are directly obtained from the moments of $\nu(g)$. By defining $\langle A^{n} \rangle$ as
the $n$-th moment of $\mathcal{P}_{\beta} (A)$, it is straightforward
to show that the first and second moments  of the amplitude are given by
\begin{equation}
\langle A \rangle = |E_{*}(\beta)|
\end{equation}
and
\begin{equation}
\langle A^2 \rangle = |E_{*}(\beta)|^2 ( 1 + \Delta_{\nu}^2 ),
\end{equation}
respectively. Hence, the contrast of the speckle pattern, which is the relative standard
deviation of $A$, is solely
determined by the standard deviation $\Delta_{\nu}$ of the number of scatterers in the medium, i.e.,
\begin{equation}
  \frac{\sqrt{\langle A^2 \rangle - \langle A \rangle^2}}{\langle A \rangle} = \Delta_{\nu}.
  \label{hds}
\end{equation}  
Clearly, strong fluctuations in the number of scatterers lead to pronounced fluctuations in the intensity
across the screen. Equations (\ref{gdew1}) and (\ref{hds}) hold for an arbitrary distribution $\nu(g)$.

\begin{figure}
  \begin{center}
    \includegraphics[scale=1.0]{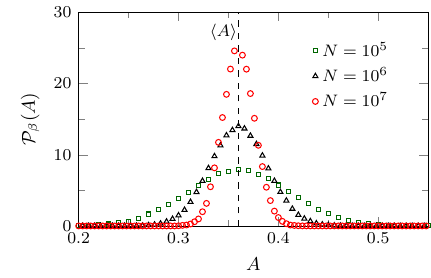} 
    \caption{Numerical results for the distribution of the amplitude for
      $\beta=\pi/2$ in the regime of weak fluctuations in the number of scatterers.
      The biased random phases are drawn
      from the bimodal distribution of Eq. (\ref{ggsj}), with $q=0.7$ and $\phi_0=\pi/4$, while the
      number of scatterers $k$ follows a Poisson distribution with average $a=\sqrt{N}$. The numerical results are generated from $10^5$ realizations of Eq. (\ref{hopi1}) with different $N$.
      The dashed vertical line marks the average amplitude $\langle A \rangle = |E_{*}(\beta)|$ for $N \rightarrow \infty$ (see Eq. (\ref{jq})).
}
\label{bias1}
\end{center}
\end{figure}

In the regime of weak fluctuations in the number of scatterers, we have that $\nu(g) = \delta(g-1)$, and Eq. (\ref{gdew1}) leads to
\begin{equation}
  \mathcal{P}_{\beta} (A) = \delta \left( A - |E_{*}(\beta)| \right).
  \label{jq}
\end{equation}  
Hence, if the tail of $p_k$ decays sufficiently fast, the distribution of the amplitude becomes
peaked at its average value $|E_{*}(\beta)|$. 
Using the explicit form of $E_{*}(\beta)$, Eq. (\ref{jjaa1}), we can rewrite $\mathcal{P}_{\beta} (A)$ as follows
\begin{equation}
  \mathcal{P}_{\beta} (A) = \delta\left[A - \sqrt{ \left\langle  \sin \phi  \right\rangle_{\phi}^2 + \left\langle  \cos \phi  \right\rangle_{\phi}^2}
    \, \frac{2E_0 }{\beta }  \sin{\left( \frac{\beta}{2}  \right)}  \right].
  \label{hhzzp}
\end{equation}  
This is the sinc function multiplied by a factor that depends on the phase distribution. Thus, for biased
distributions $\Omega(\phi)$, despite phases being generally random variables, the amplitude 
equals its average value as long as $\Delta_\nu=0$. Figure \ref{bias1} shows numerical histograms of the amplitude obtained
from Eq. (\ref{hopi1}) for different sizes $N$ in the regime of weak fluctuations in the number of scatterers.
Clearly, as $N$ increases, the histograms in figure (\ref{hopi1}) become sharply peaked at $\langle A \rangle = |E_{*}(\beta)|$, confirming
Eq. (\ref{hhzzp}). When all phases are equal to a constant, Eq. (\ref{hhzzp})
reduces to the usual sinc function.

\begin{figure}
  \begin{center}
    \includegraphics[scale=1.0]{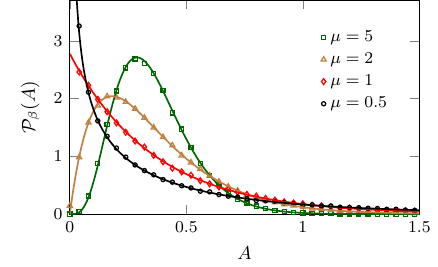} 
    \caption{Comparison between the analytic expression for the distribution of the amplitude, Eq. (\ref{hdpo}) (solid lines), and numerical results (symbols) for
      $\beta=\pi/2$ and biased random phases. The rescaled number of scatterers $\nu(g)$ follows a gamma distribution with variance $1/\mu$ (see Eq. (\ref{gfdsw})).  
      The phases are drawn from the  bimodal distribution of Eq. (\ref{ggsj}), with $q=0.7$ and $\phi_0=\pi/4$.
      The numerical results (different symbols) are obtained from $2 \times 10^5$ realizations of the model with $N=10^7$, where the number
      of scatterers is drawn from a negative binomial distribution with mean $a=\sqrt{N}$ (see Eq. (\ref{fsdp})).
}
\label{biased}
\end{center}
\end{figure}

In the regime of strong fluctuations in the number of scatterers, we combine Eqs. (\ref{gfdsw}) and (\ref{gdew1}) and obtain the explicit formula
\begin{equation}
  \mathcal{P} _{\beta} (A) = \frac{\mu^\mu}{|E_{*}(\beta)|^\mu \Gamma(\mu) } \, A^{\mu-1} e^{- \frac{\mu A}{  |E_{*}(\beta)| }}.
  \label{hdpo}
\end{equation}
Figure \ref{biased} compares Eq. (\ref{hdpo}) with histograms of the amplitude generated from numerical results for finite $N$. The latter are obtained
from Eq. (\ref{hopi1}) for several realizations of the model. Figure \ref{biased} demonstrates the exactness of Eq. (\ref{hdpo}) in depicting the
amplitude distribution in cases involving biased random phases and strong fluctuations in the number of scatterers.

\begin{figure}
  \begin{center}
    \includegraphics[scale=1.0]{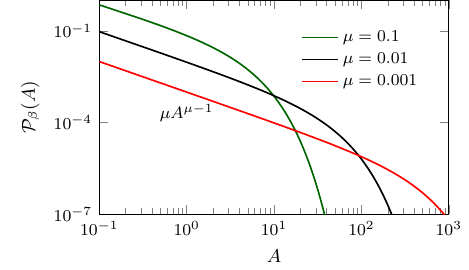} 
    \caption{Tails of the distribution of the amplitude for biased random phases and $\beta=\pi/2$. These results, obtained
      from Eq. (\ref{hdpo}), are shown in logarithmic scale. The rescaled number of scatterers follows a Gamma distribution $\nu(g)$ with variance $1/\mu$ (see Eq. (\ref{gfdsw})). 
      The phases are drawn from the  bimodal distribution of Eq. (\ref{ggsj}), with $q=0.7$ and $\phi_0=\pi/4$. 
}
\label{biased1}
\end{center}
\end{figure}

Equation (\ref{hdpo}) reveals two interesting consequences of the strong  fluctuations in the number of scatterers. The first
one concerns the dramatic change in $\mathcal{P} _{\beta} (A \rightarrow 0)$ as a function of the variance $\Delta_{\nu}^2= 1/\mu$. For $\mu > 1$, we obtain
$\lim_{A \rightarrow 0} \mathcal{P} _{\beta} (A)=0$, while the amplitude distribution diverges as a power-law $\mathcal{P} _{\beta} (A) \propto A^{\mu-1}$ ($A \rightarrow 0$) for $\mu < 1$.
For $\mu=1$,  $\lim_{A \rightarrow 0} \mathcal{P} _{\beta} (A)$ converges to a finite value. The second interesting feature concerns
the  behavior of $\mathcal{P} _{\beta} (A)$ for large amplitudes, as illustrated in figure \ref{biased1}.
For large $A$, $\mathcal{P} _{\beta} (A)$ exhibits a power-law tail $A^{\mu-1}$ ($\mu < 1$) that
extends up to a threshold of $\mathcal{O}( |E_{*}(\beta)|/\mu)$. For $A \gtrsim |E_{*}(\beta)|/\mu  $, the exponential factor in Eq. (\ref{hdpo}) becomes
important and suppresses the power-law decay. The value of the threshold that separates the power-law and exponential regimes
diverges as $\mu \rightarrow 0$, which highlights the appearance of huge 
fluctuations in the amplitude. In summary, for $\mu < 1$, although the probability to observe a vanishing
amplitude at a given point $\beta$ on the screen is very large, there is a significant probability
to observe an extremely large amplitude in comparison to $\langle A \rangle$. The occurrence of such
rare events is enhanced by increasing the fluctuations in the number of scatterers.

%%%%%%%%%%%%%%%%%%%%%%%%%%%%%%%%%%%%%%%%%%%%%%%%%%%%%%
%%%%%%%%%%%%%%%%%%%%%%%%%%%%%%%%%%%%%%%%%%%%%%%%%%%%%%%%%%%%%%%%%%%

\subsection{Unbiased phases}

Here we present the analytic results for phase distributions $\Omega(\phi)$ that satisfy the constraint $\langle e^{i \phi} \rangle_\phi =  0$. 
In this case, the distribution of the complex field $E(\beta)=E_R(\beta) + i E_I(\beta)$ reads
\begin{align}
&\mathcal{W}_{\beta}(E_R,E_I) = \frac{1}{2 \pi E_{0}^2 \sigma_R \sigma_I \sqrt{1 - \rho^2}  }
\int_{0}^{\infty} \frac{d g \, \nu(g)}{g}  \nonumber \\
&\times
\exp{\left[- \frac{1}{2 g E_{0}^2 (1 - \rho^2)  }
    \left( \frac{E_R^2}{\sigma_R^2  }  + \frac{E_I^2}{\sigma_I^2  }  - \frac{2 \rho}{\sigma_R \sigma_I} E_R E_I  \right)     \right]  },
\label{hhllp}
\end{align}
where $\sigma_R$, $\sigma_I$, and $\rho$ are given by
\begin{eqnarray}
\sigma_R^2 = \frac{1}{2} + \frac{1}{4 \beta} \left\langle \sin{(2 \beta + 2 \phi)} - \sin{(2 \phi)}     \right\rangle_{\phi}, \label{j1} \\
\sigma_I^2 =   \frac{1}{2} - \frac{1}{4 \beta} \left\langle \sin{(2 \beta + 2 \phi)} - \sin{(2 \phi)}     \right\rangle_{\phi},   \label{j2} \\
\rho = \frac{1}{4 \beta \sigma_R \sigma_I} \left\langle  \cos{(2 \phi)} - \cos{(2 \beta + 2 \phi)}  \right\rangle_{\phi}. \label{j3}
\end{eqnarray}  
The above parameters thus depend on the phase distribution $\Omega(\phi)$ as well
as on the position $\beta$ along the screen. In the regime of weak fluctuations in the number of scatterers ($\Delta_{\nu} = 0$), Eq. (\ref{hhllp})
reduces to a bivariate Gaussian distribution, which is a direct consequence of the central
limit theorem as applied to a random walk in two dimensions \cite{Goodman}. In the regime of strong fluctuations in the number of scatterers ($\Delta_{\nu} > 0$), the central limit
theorem breaks down and one has to specify $\nu(g)$ to determine $\mathcal{W}_{\beta}(E_R,E_I)$.

Equation (\ref{hhllp}) yields the analytic expression for the distribution of the amplitude  
\begin{equation}
  \mathcal{P} _{\beta} (A) = \int_{0}^{\infty} \frac{d g \, \nu(g)}{g} \mathcal{C}_{\beta}(A|g),
  \label{hopiaq}
\end{equation}  
where
\begin{align}
\mathcal{C}_{\beta}(A|g) &=  \frac{A}{ E_{0}^2 \sigma_R \sigma_I \sqrt{1 - \rho^2}  }
\exp{\left[ - \frac{  A^2  }{4 g E_{0}^2 (1 - \rho^2)\sigma_R^2 \sigma_I^2  }   \right]} \nonumber \\
& \times I_0 \left[ \frac{ \sqrt{(\sigma_R^2 - \sigma_I^2)^2 + 4 \sigma_R^2 \sigma_I^2 \rho^2} }{ 4 g E_{0}^2 (1 - \rho^2)\sigma_R^2 \sigma_I^2  } A^2  \right],
\label{hopiaq1}
\end{align}  
with $I_0(x)$ denoting the modified Bessel function of the first kind. Equations (\ref{hhllp}) and (\ref{hopiaq}) hold when both $N$ and $a$ are infinitely
large, yet the density of scatterers fulfills $D=a/N \rightarrow 0$. In appendix \ref{calc}, we cover all the specific steps involved
in the derivation of Eqs. (\ref{hhllp}) and (\ref{hopiaq}).

\begin{figure}[H]
  \begin{center}
    \includegraphics[scale=1.0]{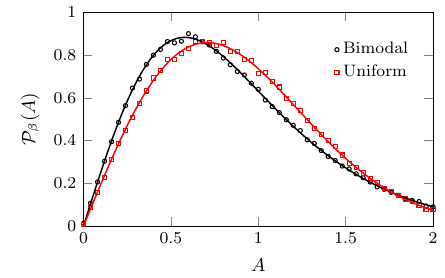} 
    \caption{Comparison between Eq. (\ref{gdlm}) (solid lines) and
      numerical histograms (symbols) for $\beta=\pi/2$ and weak fluctuations in the number of scatterers. 
      The unbiased random phases follow an uniform distribution in the interval $[0,2 \pi)$ or they
        are sampled from the bimodal distribution of Eq. (\ref{ggsj}), with $q=0.5$ and $\phi_0=\pi/4$. The
        number of scatterers $k$ follows a Poisson distribution with average $a=\sqrt{N}$. The numerical results are generated from $2 \times 10^5$ realizations
        of Eq. (\ref{hopi1}) with $N=10^7$.
}
\label{unbiasedWeak}
\end{center}
\end{figure}

Equation (\ref{hopiaq}) is one of the main findings of our work, since it provides the amplitude distribution for
unbiased random phases and any pair of distributions $\Omega(\phi)$ and $\nu(g)$, generalizing classic results \cite{Goodman,Jakeman1978,Jakeman1980}
in the theory of speckle patterns. Let us analyze a few limiting cases of Eq. (\ref{hopiaq}). In the regime of weak fluctuations
in the number of scatterers ($\nu(g) = \delta(g-1)$), we get
\begin{align}
  \mathcal{P} _{\beta} (A) &=  \frac{A}{ E_{0}^2 \sigma_R \sigma_I \sqrt{1 - \rho^2}  }
\exp{\left[ - \frac{ A^2 }{4  E_{0}^2 (1 - \rho^2)\sigma_R^2 \sigma_I^2  }  \right]} \nonumber \\
&\times I_0 \left[ \frac{ \sqrt{(\sigma_R^2 - \sigma_I^2)^2 + 4 \sigma_R^2 \sigma_I^2 \rho^2} }{ 4  E_{0}^2 (1 - \rho^2)\sigma_R^2 \sigma_I^2  } A^2  \right].
\label{gdlm}
\end{align}  
Equation (\ref{gdlm}) depicts a novel amplitude distribution that applies to any phase distribution, as long as it satisfies the soft constraint $\langle e^{i \phi} \rangle_{\phi}=0$.
Since the argument of the Bessel function in the above equation is proportional to $A^2$, Eq. (\ref{gdlm}) is distinct from a Rice distribution \cite{Goodman}. The
latter describes the statistics of the amplitude produced by a finite constant field plus a large number
of small random fields with uniformly distributed phases \cite{Goodman}. When the phases are continuous random variables sampled from
an uniform distribution in the interval $[0,2 \pi)$, Eqs. (\ref{j1}-\ref{j3}) result in
\begin{equation}
  \sigma_R^2 = \sigma_I^2 = \frac{1}{2}, \quad \rho=0,
  \label{juio}
\end{equation}  
and Eq. (\ref{gdlm}) simplifies to the well-known Rayleigh distribution
\begin{equation}
\mathcal{P} (A) = \frac{2 A}{E_0^2} \exp{\left( -\frac{A^2}{E_0^2} \right)}.
\label{Ray}
\end{equation} 
Thus, Eq. (\ref{gdlm}) essentially generalizes the Rayleigh distribution to
non-uniform phase distributions.
Figure \ref{unbiasedWeak} compares Eq. (\ref{gdlm}) with numerical histograms of the amplitude for
two distinct distributions of unbiased phases in the regime of weak fluctuations in the number of scatterers. The numerical results
are fully consistent with our analytic expression.

For random phases with an uniform distribution, we can substitute Eqs. (\ref{juio}) in Eq. (\ref{hopiaq1}), resulting in
\begin{equation}
  \mathcal{P} (A) =  \frac{2 A}{E_0^2} \int_{0}^{\infty} \frac{d g \, \nu(g)}{g} e^{- \frac{A^2}{g E_0^2} }.
  \label{loopa}
\end{equation}  
The distribution $\mathcal{P} (A)$ above is independent of the position $\beta$ along the screen, regardless the specific form
of $\nu(g)$.
When  fluctuations in the number of scatterers are large ($\Delta_\nu > 0$), the universality associated with
the central limit theorem breaks down. As a result, the behavior of $\mathcal{P} (A)$ depends on the distribution $\nu(g)$. For example, when
$\nu(g)$ follows a gamma distribution, we can substitute Eq. (\ref{gfdsw})
into Eq. (\ref{loopa}), integrate over the variable $g$, and arrive at the so-called $K$-distribution  \cite{Jakeman1978,Jakeman1980}
\begin{equation}
  \mathcal{P} (A) = \frac{4 \mu^{\frac{\mu+1}{2}}}{E_0^{\mu+1} \Gamma(\mu)} A^{\mu} K_{\mu-1}\left( \frac{2 \sqrt{\mu}}{E_0} A \right),
  \label{jusw}
\end{equation}  
with $K_{\mu}(x)$ representing a modified Bessel function of the second kind.

Let us now turn our attention to the general Eq. (\ref{hopiaq}). 
Depending on the specific form of $\nu(g)$, the integral
in Eq. (\ref{hopiaq}) has no analytic solution and one cannot derive a closed-form expression for $\mathcal{P} _{\beta} (A)$. However, when $\nu(g)$
is given by the gamma distribution, Eq. (\ref{gfdsw}), we show
in appendix \ref{calc1} that, for any positive integer $\mu=n=1,2,\dots$, the amplitude distribution can be
calculated from the identity
\begin{equation}
  \mathcal{P} _{\beta} (A) = \frac{ A}{E_0^2 \sigma_R \sigma_I \sqrt{1-\rho^2}} \frac{(-1)^{n-1} n^n}{(n-1)!}
  \frac{\partial^{n-1} H(u)}{\partial u^{n-1}} \Bigg{|}_{u=n},
  \label{der}
\end{equation}  
where $H(u)$ is given by
\begin{equation}
  H(u) = 2 I_0\left[ \frac{A \, f(\omega) \sqrt{u}}{E_0 \sigma_R\sigma_I \sqrt{1-\rho^2} }  \right] K_0\left[ \frac{A \, g(\omega) \sqrt{u}}{E_0 \sigma_R\sigma_I \sqrt{1-\rho^2} }  \right].
  \label{hspopo}
\end{equation}  
The functions $f(\omega)$ and $g(\omega)$ are defined as
\begin{eqnarray}
    f(\omega)&=& 
\begin{cases}
    \sin{\omega}, & \text{if }  0 \leq \omega \leq \pi/4 \\
    \cos{\omega}, &  \text{if }  \pi/4 < \omega \leq \pi/2,
\end{cases} \\
    g(\omega)&=& 
\begin{cases}
    \cos{\omega}, & \text{if }  0 \leq \omega \leq \pi/4 \\
    \sin{\omega}, &  \text{if }  \pi/4 < \omega \leq \pi/2,
\end{cases}
\end{eqnarray}
where $\omega \in [0,\pi/2]$ is determined by $\Omega(\phi)$ as follows
\begin{equation}
  \omega = \frac{1}{2} \sin^{-1}{\left( \sqrt{(\sigma_R^2 - \sigma_I^2)^2 + 4 \sigma_R^2 \sigma_I^2 \rho^2     } \right)}.
  \label{kkop}
\end{equation}
Equation (\ref{der}) provides a practical way to obtain closed-form analytic expressions for $\mathcal{P} _{\beta} (A)$ when $\nu(g)$ is
given by Eq. (\ref{gfdsw}), with integer $\mu=n>0$, and $\Omega(\phi)$ is an arbitrary distribution. Thus, Eq. (\ref{der}) generalizes
the $K$-distribution of the amplitude, which is specific to uniform random phases, to any distribution of unbiased phases.
The analytic expressions for $\mathcal{P} _{\beta} (A)$ when $\mu=1$ and $\mu=2$ are, respectively, given by
\begin{equation}
  \mathcal{P} _{\beta} (A) =  \frac{2 A}{E_0 \gamma}
  I_0\left( \frac{A f(\omega) }{\gamma }  \right) K_0\left( \frac{A g(\omega)}{\gamma}  \right)
  \label{b1}
\end{equation}
and
\begin{align}
  \mathcal{P} _{\beta} (A) &=  \frac{2 \sqrt{2} A^2}{E_0 \gamma^2}
  \Bigg{[} g(\omega)  I_0 \left( \frac{\sqrt{2} A f(\omega) }{\gamma }  \right)  K_1 \left( \frac{\sqrt{2} A g(\omega)}{\gamma } \right) \nonumber \\
    &- f(\omega)  I_1 \left( \frac{\sqrt{2} A f(\omega) }{\gamma }  \right)  K_0 \left( \frac{\sqrt{2} A g(\omega)}{\gamma } \right) \Bigg{]},
  \label{b2}
\end{align}
with
\begin{equation}
\gamma = E_0 \sigma_R\sigma_I \sqrt{1-\rho^2}.
\end{equation}  
In appendix \ref{calc1}, we explain how to derive Eq. (\ref{der}).

\begin{figure}[H]
  \begin{center}
    \includegraphics[scale=1.0]{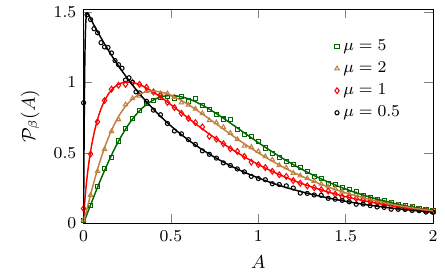} 
    \caption{Comparison between the theoretical results (solid lines) for the amplitude distribution and numerical results (symbols) for
      $\beta=\pi/2$ and unbiased random phases. The rescaled number of scatterers $\nu(g)$ follows a gamma distribution with variance $1/\mu$ (see Eq. (\ref{gfdsw})). 
      The phases are drawn from the bimodal distribution of Eq. (\ref{ggsj}), with $q=0.5$
      and $\phi_0=\pi/4$. The numerical results (different symbols) are obtained from $2 \times 10^5$ realizations of the model with $N=10^7$, where the number
      of scatterers is drawn from a negative binomial distribution with mean $a=\sqrt{N}$ (see Eq. (\ref{fsdp})). 
}
\label{unbiased}
\end{center}
\end{figure}

In fig. \ref{unbiased}, we compare our theoretical results with numerical histograms obtained from Eq. (\ref{hopi1}) in the regime of strong
fluctuations in the number of scatterers. The solid curves for $\mu=1$ and $\mu=2$ are obtained, respectively, from Eqs. (\ref{b1}) and (\ref{b2}), while
the theoretical results for $\mu <1$ are derived by numerically solving the integral in Eq. (\ref{hopiaq}). The remarkable
consistency between our analytic findings and numerical simulations confirm the exactness of our theory.

There are two distinctive features of $\mathcal{P}_{\beta}(A)$ when the phases are unbiased random variables
and $\nu(g)$ follows Eq. (\ref{gfdsw}). First, as shown in
figure \ref{unbiased}, the distribution $\mathcal{P}_{\beta}(A)$ goes to zero at $A=0$ for any value of $\mu$, in stark contrast to the behavior of $\mathcal{P}_{\beta}(A)$
for biased random phases (see Eq. (\ref{hdpo})). The complex field $E$ is a sum of a large number of independent random variables.
  For unbiased phases, the fraction of samples or realizations with $E$ {\it exactly equal to zero} decreases exponentially with $N$ \cite{Touchette2009}. Therefore, in the
  limit $N \rightarrow \infty$, the fraction of samples with $A=0$ goes
  to zero, regardless the value of $\mu$ or the choice of the phase distribution. For biased phases, the statistics of $E$ is controlled
  by the distribution of the rescaled number $g=k/a$ of scatterers. In the limit $a \rightarrow \infty$, samples that
  have a small number of scatterers in comparison to the mean $a$ contribute with $E=0$ to the statistics of $E$.
  Therefore, the fact that  $\mathcal{P}_{\beta}(A \rightarrow 0)$ is nonzero for biased phases is a consequence of the finite fraction
  of samples with $g = k/a \rightarrow 0$.

\begin{figure}[H]
  \begin{center}
    \includegraphics[scale=1.0]{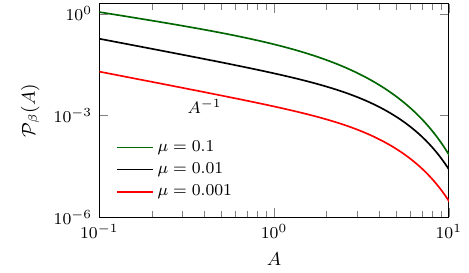} 
    \caption{Tails of the distribution of the amplitude for unbiased random phases and  $\beta=\pi/2$. These results, obtained
      from Eq. (\ref{hdpo}), are shown in logarithmic scale. The rescaled number of scatterers follows a Gamma distribution $\nu(g)$ with variance $1/\mu$ (see Eq. (\ref{gfdsw})).
      The phases are drawn from the  bimodal distribution of Eq. (\ref{ggsj}), with $q=0.5$ and $\phi_0=\pi/4$.
}
\label{unbiased1}
\end{center}
\end{figure}

The second interesting property concerns the right tail of the amplitude  distribution. In figure
\ref{unbiased1}, we plot $\mathcal{P}_{\beta}(A)$ for large $A$ in the case of unbiased phases. For small $\mu$ and, consequently, strong fluctuations in
the number of scatterers, the distribution $\mathcal{P}_{\beta}(A)$ exhibits once more a power-law tail up to a certain threshold, above which $\mathcal{P}_{\beta}(A)$ decays exponentially fast. Although
this behavior is similar to the biased case, there is a key difference: the threshold value in figure \ref{unbiased1}
remains approximately independent of $\mu$. For this reason, the probability of observing large fluctuations of the amplitude 
for unbiased phases is much smaller than for biased phases, for which the threshold value diverges as $\mu \rightarrow 0$ (see figure \ref{biased1}).
%

%%%%%%%%%%%%%%%%%%%%%%%%%%%%%%%%%%%%%%%%%%%%%%%%%%%%%%%%%%%%%%%%%%%%%%%%%%%%%%%%%%%%%
%%%%%%%%%%%%%%%%%%%%%%%%%%%%%%%%%%%%%%%%%%%%%%%%%%%%%%%%%%%%%%%%%%%%%%%%%%%%%%%%%%%%%%%%%%%%%%

\section{Final remarks}
\label{Final}

In this work, we have developed a comprehensive theory of speckle patterns based on the random
walk model. In this model, the resultant speckle field is the superposition of a stochastic number of
partial waves, each with a random phase. The distribution of the number of scatterers as well as the
distribution of phases are inputs of the model. By distinguishing between biased and unbiased random phases, we have
derived general equations for the amplitude distribution of the speckle field in the
limit of an infinitely large mean number of scatterers.
Once the phase distribution and the distribution
of the number of scatterers are specified, our main findings, Eqs. (\ref{gdew1}) and (\ref{hopiaq}), lead to closed-form
analytic expressions that encompass a broad range of situations.

Two families of analytic results for unbiased random phases deserve
a special attention. First, when the number of scatterers is nonrandom, Eq. (\ref{hopiaq}) yields a novel
form of the amplitude distribution which is the natural
generalization of the Rayleigh law to nonuniform random phases.
Second, when the number of scatterers is drawn from a negative binomial distribution with an integer
scale parameter, Eq. (\ref{hopiaq}) gives rise to generalizations of the $K$-distribution
to nonuniform phases. We have confirmed the exactness of these results by comparing
them with numerical simulations for a bimodal phase distribution.

Interestingly, we have shown that the behavior of the amplitude distribution $\mathcal{P}_{\beta}(A)$ for large $A$ is qualitatively
distinct in the cases of biased and unbiased random phases.
In both situations, $\mathcal{P}_{\beta}(A)$ displays a power-law decay up to
a certain threshold, above which it decays exponentially fast (see figures \ref{biased1} and \ref{unbiased1}).
The difference appears in the threshold behavior as a function of the variance $\Delta_{\nu}^2$
of the number of scatterers.
While for biased random phases this threshold increases as a function of $\Delta_{\nu}^2$, it remains approximately constant for unbiased random phases.
This implies that a certain degree of phase coherence favors the generation of extremely large amplitudes, in line with
previous works \cite{Bonatto2020,daSilva2020}. In this context, it would be interesting
to quantify the impact of fluctuations of the number of scatterers
in the formation of rogue intensity waves \cite{Bonatto2020,Buarque2023}.

We expect that our analytic findings can be experimentally tested using a spatial light
modulator (SLM). This device replicates the scattering from a rough surface through
the phase modulation of an incident wave. When the SLM is illuminated by a laser beam, each pixel's diffraction on the SLM mask generates a
partial wave with a specific phase. In principle, it is possible to imprint any sequence of random phases on the SLM \cite{Bromberg2014,Bender2019,Bonatto2020}, making it
an ideal platform for testing our analytic predictions in the case of nonuniform phase distributions. However, an important
drawback in this experimental setup is the limited number of pixels.
Our main results for the amplitude distribution, Eqs. (\ref{gdew1}) and (\ref{hopiaq}), are valid in the regime of low density of scatterers. This
implies that both the sample size $N$ (total number of pixels in a transversal direction) and
the mean number of scatterers $a$ (mean number of active pixels) are both very large, but the density $a/N$ goes to zero. In general, approaching
this limit in SLM experiments may be a difficult task.

It would be interesting to compare our analytic findings for the amplitude distribution with results obtained from the numerical solutions of the Maxwell equations
for large assemblies of scatterers distributed in space with a controllable density \cite{Pattelli2018}. In this way, one could
test whether the results obtained from the random walk model serve as approximations to other
relevant scenarios, such as in the study of speckle patterns formed by three-dimensional samples \cite{2019Skipetrov} or in the near field regime.
Another interesting research line is to investigate the connection between the parameters of our phenomenological random walk
model and the structural features characterizing models of interacting scattering particles, such as particle size and spatial correlations
in the positions of the scatterers \cite{Almasian2017}. In this context, one expects that biased phase distributions are relevant to model
the effects arising from spatial correlations.

To summarize, our analytic results expand the scope of the random walk model
of speckle patterns and open the perspective to systematically investigate the role of
nonuniform phase distributions in a wide range of problems
involving the linear superposition of random waves \cite{Dudley2019}. Our analytic techniques and results should be particularly
useful to study non-isotropic random walks in finite dimensions \cite{Jakeman1987,Barber1993,Gregorio2012}.

\begin{acknowledgments}

The authors thank R.~R.~B.~Correia for useful comments. F. L. M. thanks CNPq/Brazil for financial support.
  
\end{acknowledgments}

%%%%%%%%%%%%%%%%%%%%%%%%%%%%%%%%%%%%%%%%%%%%%%%%%%%%%%%%%%%%%%%%%%%%%%%%%%%%%%%%%%
%%%%%%%%%%%%%%%%%%%%%%%%%%%%%%%%%%%%%%%%%%%%%%%%%%%%%%%%%%%%%%%%%%%%%%%%%%%%%%%%%%%
%%%%%%%%%%%%%%%%%%%%%%%%%%%%%%%%%%%%%%%%%%%%%%%%%%%%%%%%%%%%%%%%%%%%%%%%%%%%%%%%%%%

\appendix

\section{Derivation of the amplitude distribution}
\label{calc}

In this appendix we explain how to obtain the analytic results for the amplitude distribution. The first step
is to calculate the characteristic function $\mathcal{G}_{\beta}(u,v)$ of the joint distribution $\mathcal{W}_{\beta}(E_R,E_I)$ of the real
and imaginary parts of the speckle field $E(\beta)=E_R(\beta) + i E_I(\beta)$. The function $\mathcal{G}_{\beta}(u,v)$ is defined as
\begin{align}
  \mathcal{G}_{\beta}(u,v) &= \Bigg{\langle} \exp{\Big{[} -i u E_0 \sum\limits_{j=0}^{N-1} x_j {\rm Re} \mathcal{E}_j(\phi_j) \Big{]}} \nonumber \\
 &\times \exp{\Big{[} -i v E_0 \sum\limits_{j=0}^{N-1} x_j {\rm Im} \mathcal{E}_j(\phi_j) \Big{]} }  \Bigg{\rangle}_{\boldsymbol{x},\boldsymbol{\phi} },
  \label{jjuq}
\end{align}  
with
\begin{equation}
  \mathcal{E}_j(\phi_j) = e^{\frac{i j \beta }{N-1} + i \phi_j }.
\end{equation}
The symbols $\langle \dots \rangle_{\boldsymbol{x}}$ and $\langle \dots \rangle_{\boldsymbol{\phi}}$ denote, respectively, the average over
$x_{0},\dots,x_{N-1}$ and $\phi_0,\dots,\phi_{N-1}$. The Fourier transform of $\mathcal{G}_{\beta}(u,v)$ yields $\mathcal{W}_{\beta}(E_R,E_I)$, namely

\begin{equation}
  \mathcal{W}_{\beta}(E_R,E_I) = \int_{-\infty}^{\infty} \frac{du dv}{4 \pi^2} e^{i u E_R + i v E_I} \mathcal{G}_{\beta}(u,v).
  \label{popoq}
\end{equation}
In order to calculate the average over $x_{0},\dots,x_{N-1}$ with the joint distribution (\ref{rigt1}), we need to factorize the conditional probability $P(\boldsymbol{x} |k )$, defined
in Eq. (\ref{rigt}), as a product over the index $j=0,\dots,N-1$ that identifies the scatterers. This is achieved by using the following integral representation of the Kronecker-$\delta$ 
\begin{equation}
\delta_{k,\sum_{j=0}^{N-1}x_j } = \int_{0}^{2 \pi} \frac{d t}{2 \pi} \exp{\left[ i t \left( k -  \sum\limits_{j=0}^{N-1}x_j \right) \right]},
\end{equation}  
which allows us to rewrite $P(\boldsymbol{x} |k )$ as 
\begin{align}
 P(\boldsymbol{x} |k ) &= \frac{1}{\mathcal{N}_k} \int_{0}^{2 \pi} \frac{d t}{2 \pi} e^{i t k}   \nonumber  \\
 &\times \prod_{j=0}^{N-1} e^{- i t x_j } \left[ \frac{k}{N} \delta_{x_j,1} + \left(1 - \frac{k}{N}  \right) \delta_{x_j,0} \right].
\end{align}  
Combining the above expression with Eq. (\ref{rigt1}) and inserting the resulting form of $p(\boldsymbol{x})$  in Eq. (\ref{jjuq}), we compute
the average over $\boldsymbol{x}$ and obtain
\begin{align}
 &\mathcal{G}_{\beta}(u,v) = \sum_{k=0}^N \frac{p_k}{\mathcal{N}_k} \int_{0}^{2 \pi} \frac{d t}{2 \pi} e^{i t k}  \nonumber \\
  &\times \exp{\left[\sum_{j=0}^{N-1} \ln \left( 1 + \frac{k}{N} \left[e^{- i t}  T_j(u,v)   - 1    \right]\right) \right]},
  \label{mqpy}
\end{align}  
where we defined
\begin{equation}
T_j(u,v) = \left\langle e^{-i u E_0 {\rm Re} \mathcal{E}_j(\phi) - i v E_0 {\rm Im} \mathcal{E}_j(\phi) } \right\rangle_{\phi}.
\end{equation}  
The average $\langle f(\phi) \rangle_{\phi}$ of an arbitrary function $f(\phi)$ of a single phase $\phi$ is defined in Eq. (\ref{skcm}).

In the limit $N \rightarrow \infty$, we can expand the logarithm in Eq. (\ref{mqpy}) up to $\mathcal{O}(1/N)$ and replace the sum over $j=0,\dots,N-1$ by an
integral over $y \in [0,1]$
\begin{align}
 &\mathcal{G}_{\beta}(u,v) = \sum_{k=0}^\infty \frac{p_k}{\mathcal{N}^{(\infty)}_k} \int_{0}^{2 \pi} \frac{d t}{2 \pi} e^{-k + i t k}  \nonumber \\
  &\times \exp{\left[  k e^{- i t} \int_{0}^{1} dy \left\langle e^{-i u  {\rm Re} \mathcal{E}_{\phi}(y) - i v  {\rm Im} \mathcal{E}_{\phi}(y) } \right\rangle_{\phi} \right]},
  \label{trew}
\end{align}  
where 
\begin{equation}
\mathcal{E}_{\phi}(y) = E_0 e^{i \beta y + i \phi},  
\end{equation}  
and $\mathcal{N}^{(\infty)}_k = \frac{e^{-k} k^k }{k!}$ is the analytic expression for the normalization
factor $\mathcal{N}_k$ when $N \rightarrow \infty$.
By representing the second line of Eq. (\ref{trew}) as a power-series, we can integrate over $t$ and arrive at the expression
for the characteristic function in the limit $N \rightarrow \infty$, 
\begin{equation}
  \mathcal{G}_{\beta}(u,v) = \sum_{k=0}^\infty p_k e^{\frac{k}{a} Z_a(u,v) } ,
  \label{hyeda}
\end{equation}  
with
\begin{equation}
  Z_{a}(u,v) = a  \ln{ \left[   \int_{0}^{1} dy \left\langle e^{-i u {\rm Re} \mathcal{E}_{\phi}(y) - i v {\rm Im} \mathcal{E}_{\phi}(y) } \right\rangle_{\phi}   \right]}.
  \label{ute11}
\end{equation}  
Equations (\ref{hyeda}) and (\ref{ute11}) are valid as $N \rightarrow \infty$ while keeping $a$ finite. 
The next step is to perform the limit $a \rightarrow \infty $ in the above equations.
Since we first take the limit $N \rightarrow \infty$ followed by $a \rightarrow \infty$, the outcome of this order of limits
is an analytic expression for the characteristic function $\mathcal{G}_{\beta}(E_R,E_I)$ in the low density regime, i.e., when $D=\frac{a}{N} \rightarrow 0$.
To perform the limit $a \rightarrow \infty$, we need to distinguish between biased and unbiased phases.

%%%%%%%%%%%%%%%%%%%%%%%%%%%%%%%%%%%%%%%%%%%%%%%%%%%%%%%%%%%%%%%%%%
%%%%%%%%%%%%%%%%%%%%%%%%%%%%%%%%%%%%%%%%%%%%%%%%%%%%%%%%%%%%%%%%%%%

\subsection{Biased phases}

Let us consider phase distributions $\Omega(\phi)$ that fulfill the condition
\begin{equation}
  \langle e^{i \phi} \rangle_{\phi} \neq 0.
  \label{huwa}
\end{equation}  
In this case, the random variable $\mathcal{E}_\phi (y)$ fluctuates around
an average orientation in the complex plane. The most representative example of this family of distributions is when all phases are equal to a constant.

In order to perform the limit $\lim_{a \rightarrow \infty} Z_{a}(u,v)$ for this class of phase distributions, we
have to rescale the amplitudes of the individual waves as $E_0 \rightarrow E_0/a$, and Eq. (\ref{ute11}) takes
the form
\begin{align}
 Z_{a}(u,v) = a \ln{ \left[  \int_{0}^{1} dy \left\langle e^{-\frac{i u}{a} {\rm Re} \mathcal{E}_{\phi}(y) - \frac{i v}{a} {\rm Im} \mathcal{E}_{\phi}(y) }  \right\rangle_{\phi}   \right] }.
\end{align}
In the limit $a \rightarrow \infty$, $Z_{a}(u,v)$ converges to the expression
\begin{equation}
Z_{\infty}(u,v) = - i u  \int_{0}^{1} dy \langle {\rm Re} \mathcal{E}_{\phi}(y) \rangle_{\phi} - i v \int_{0}^{1} dy \langle {\rm Im} \mathcal{E}_{\phi}(y) \rangle_{\phi} .
\end{equation}  
Thus, by introducing the distribution $\nu(g)$ of the rescaled number of scatterers, Eq. (\ref{hgds8}),
the limit $a \rightarrow \infty$ of Eq. (\ref{hyeda}) is given by
\begin{equation}
  \mathcal{G}_{\beta}(u,v) = \int_{0}^{\infty} d g \nu(g) e^{g Z_{\infty}(u,v)}.
  \label{dwesa}
\end{equation}  
From Eq. (\ref{popoq}), we thus find the corresponding expression for the joint distribution of the speckle field
\begin{align}
\mathcal{W}_{\beta}(E_R,E_I) &= \int_{0}^{\infty} d g \nu(g) \delta \left[E_R - g \int_{0}^{1} dy \langle {\rm Re} \mathcal{E}_{\phi}(y)  \rangle_{\phi}   \right] \nonumber \\
&\times \delta \left[E_I - g  \int_{0}^{1} dy \langle {\rm Im} \mathcal{E}_{\phi}(y) \rangle_{\phi}    \right].
\end{align}  
Therefore, the fluctuations of the real and the imaginary parts of $E(\beta)$ are solely determined
by $\nu(g)$, with the amplitude $A(\beta)$ relating to $g$ as follows
\begin{equation}
  A(\beta) \, {\buildrel d \over =} \, g  |E_*(\beta)|,
  \label{huhu}
\end{equation}  
where
\begin{equation}
E_*(\beta) =  \int_{0}^{1} dy \langle {\rm Re} \mathcal{E}_{\phi}(y)  \rangle_{\phi}  + i  \int_{0}^{1} dy \langle {\rm Im} \mathcal{E}_{\phi}(y) \rangle_{\phi}.
\end{equation}
The symbol ${\buildrel d \over =}$ in Eq. (\ref{huhu}) means that both sides of the equation are equal in a distributional sense. Equation  (\ref{huhu})
immediately implies that the amplitude distribution $\mathcal{P}_{\beta}(A)$ is determined by $\nu(g)$ according to Eq. (\ref{gdew1}).

%%%%%%%%%%%%%%%%%%%%%%%%%%%%%%%%%%%%%%%%%%%%%%%%%%%%%%%%%%%%%%%%%%%%%%
%%%%%%%%%%%%%%%%%%%%%%%%%%%%%%%%%%%%%%%%%%%%%%%%%%%%%%%%%%%%%%%%%%%%%%%%%

\subsection{Unbiased phases}

Here we consider phase distributions $\Omega(\phi)$ that satisfy the constraint
\begin{equation}
  \langle e^{i \phi} \rangle_{\phi} = 0.
  \label{huwa1}
\end{equation}  
In this case, the average of the complex random variable $\mathcal{E}_\phi (y)$ is zero. The most representative example of this class of distributions is when the phases
are continuous random variables drawn from an uniform distribution in $[0,2 \pi)$.

For phase distributions that fulfill Eq. (\ref{huwa1}), we rescale the amplitude $E_0$ as $E_0 \rightarrow E_0/\sqrt{a}$, and Eq. (\ref{ute11}) assumes the form
\begin{align}
  Z_{a}(u,v) = a \ln{ \left[  \int_{0}^{1} dy \left\langle e^{-\frac{i u}{\sqrt{a}}  {\rm Re} \mathcal{E}_{\phi}(y)
        - \frac{i v}{\sqrt{a}} {\rm Im} \mathcal{E}_{\phi}(y)   }  \right\rangle_{\phi}   \right] }.
\end{align}
By expanding the right hand side of the above equation in powers of $1/\sqrt{a}$, one can show that $\lim_{a \rightarrow \infty} Z_{a}(u,v)$ is given by
the quadratic form
\begin{equation}
Z_{\infty}(u,v) = - \frac{1}{2} E_0^2 \sigma_R^2 u^2  - \frac{1}{2} E_0^2 \sigma_I^2 v^2 - u v E_0^2 \sigma_R \sigma_I \rho,
\end{equation}  
where $\sigma_R^2$, $\sigma_I^2$, and $\rho$ are defined by Eqs. (\ref{j1}-\ref{j3}). Substituting this result
in Eq. (\ref{hyeda}) and taking the limit $a \rightarrow \infty$, we obtain the characteristic function
\begin{equation}
  \mathcal{G}_{\beta}(u,v) = \int_{0}^{\infty} d g \nu(g) e^{g Z_{\infty}(u,v)}.
  \label{eewg}
\end{equation}  
Inserting Eq. (\ref{eewg}) in Eq.  (\ref{popoq}) and calculating the Gaussian integrals over $u$ and $v$, we find
the joint distribution of the complex field
\begin{align}
  &\mathcal{W}_{\beta}(E_R,E_I) = \frac{1}{2 \pi E_0^2 \sigma_R \sigma_I \sqrt{1 - \rho^2}} \int_{0}^{\infty} \frac{dg}{g} \nu(g) \nonumber \\
  &\times \exp{\left[- \frac{1}{2 g E_{0}^2 (1-\rho^2)} \left( \frac{E_R^2}{\sigma_R^2} + \frac{E_I^2}{\sigma_I^2} - \frac{2 \rho E_R E_I}{\sigma_R \sigma_I}   \right) \right]}.
  \label{rrt}
\end{align}  
When the distribution of the number of scatterers is such that $\nu(g) = \delta(g-1)$, $\mathcal{W}_{\beta}(E_R,E_I)$ follows a bivariate
Gaussian distribution, which is a consequence of the central limit theorem. When the variance of $\nu(g)$ is finite, the central
limit theorem fails and $\mathcal{W}_{\beta}(E_R,E_I)$ depends on the form of $\nu(g)$.

Equation (\ref{rrt}) enables to obtain the amplitude distribution for any $\nu(g)$. Let $G_{\beta}(S)$ be the characteristic function
of the probability density $P_{\beta}(I)$ of the  intensity $I {\buildrel d \over =} E_R^2 + E_I^2$, namely
\begin{equation}
  G_{\beta}(S) = \int_{-\infty}^{\infty} d E_R d E_I  \mathcal{W}_{\beta}(E_R,E_I) e^{-i S \left( E_R^2 + E_I^2  \right)}.
  \label{hytra}
\end{equation}  
The distribution $P_{\beta}(I)$ follows from the Fourier transform
\begin{equation}
  P_{\beta}(I) = \int_{-\infty}^{\infty} \frac{dS}{2 \pi} e^{i S I} G_{\beta}(S).
  \label{hhpp}
\end{equation}  
Substituting Eq. (\ref{rrt}) into Eq. (\ref{hytra}) and performing the Gaussian integrals over $E_R$ and $E_I$, we arrive at the expression
\begin{align}
  &G_{\beta}(S) = \frac{1}{E_0^2 \sigma_R \sigma_I \sqrt{1 - \rho^2}} \int_{0}^{\infty} \frac{dg}{g} \nu(g) \nonumber \\
  &\times \Bigg[ \left(2 i S + \frac{1}{g E_0^2 (1-\rho^2) \sigma_R^2  } \right) \left(2 i S + \frac{1}{g E_0^2 (1-\rho^2) \sigma_I^2  } \right) \nonumber \\
    &- \frac{\rho^2}{g^2 E_0^4 (1-\rho^2)^2 \sigma_R^2 \sigma_I^2  }   \Bigg]^{-\frac{1}{2}}.
  \label{plij}
\end{align}  
The final act is the calculation of the Fourier transform in Eq. (\ref{hhpp}). Combining Eqs. (\ref{plij}) and (\ref{hhpp}), we can rewrite $P_{\beta}(I)$ as follows
\begin{align}
 &P_{\beta}(I) = \frac{1}{4 \pi E_0^2 \sigma_R \sigma_I \sqrt{1 - \rho^2}} \int_{0}^{\infty} \frac{dg}{g} \nu(g) \nonumber \\
  & \times \int_{-\infty}^{\infty} \frac{dS \, e^{- i S I } }{\left(r_{+} - i S   \right)^{\frac{1}{2}} \left(r_{-} - i S   \right)^{\frac{1}{2}}  },
  \label{hytr}
\end{align}  
where we have defined 
\begin{align}
  r_{\pm} &= \frac{1}{4 g E_0^2 (1-\rho^2) \sigma_R^2 \sigma_I^2 } \nonumber \\
  &\times \left(\sigma_R^2 + \sigma_I^2 \pm \sqrt{\left( \sigma_R^2 - \sigma_I^2 \right)^2 + 4 \sigma_R^2  \sigma_I^2 \rho^2 } \right).
\end{align}  
By following \cite{gradshteyn}, we integrate over the variable $S$ in Eq. (\ref{hytr}), obtaining
\begin{align}
  P_{\beta} (I) &= \frac{1}{2 E_{0}^2 \sigma_R \sigma_I \sqrt{1 - \rho^2}  } \int_{0}^{\infty} \frac{d g \, \nu(g)}{g}
  e^{- \frac{  I  }{4 g E_{0}^2 (1 - \rho^2)\sigma_R^2 \sigma_I^2  } } \nonumber \\
  &\times I_0 \left( \frac{ \sqrt{(\sigma_R^2 - \sigma_I^2)^2 + 4 \sigma_R^2 \sigma_I^2 \rho^2} }{ 4 g E_{0}^2 (1 - \rho^2)\sigma_R^2 \sigma_I^2  } I  \right),
  \label{hopiaq11}
\end{align}  
where $I_0(x)$ is a modified Bessel function of the first kind. The above expression gives the distribution
of the intensity $I {\buildrel d \over =} A^2$. We find Eq. (\ref{hopiaq}) for the amplitude distribution $\mathcal{P}_{\beta} (A)$
by making a simple change of variables.

%%%%%%%%%%%%%%%%%%%%%%%%%%%%%%%%%%%%%%%%%%%%%%%%%%%%%%%%%%%%

\section{Amplitude distribution for integer $\mu$}
\label{calc1}

Here we explain how to obtain Eq. (\ref{der}), which leads to analytic expressions for $\mathcal{P}_{\beta}(A)$ when
$\nu(g)$ is given by the gamma distribution of Eq. (\ref{gfdsw}) with integer $\mu$. Substituting
Eq. (\ref{gfdsw}) into Eq. (\ref{hopiaq}) and setting $\mu=n \in \mathbb{Z}^{+}$, we get the general expression
\begin{align}
&\mathcal{P}_{\beta}(A) = \frac{A}{E_{0}^2 \sigma_R \sigma_I \sqrt{1 - \rho^2}} \frac{n^n}{2^{n-1} (n-1)!}
\int_{0}^{\infty} dg \, g^{n-2} e^{- \frac{n g}{2}} \nonumber \\
& \times \exp{\left(- \frac{A^2}{2 E_{0}^2 (1-\rho^2) \sigma_R^2 \sigma_I^2 g }  \right)} \nonumber \\
& \times I_0 \left( \frac{\sqrt{(\sigma_R^2 - \sigma_I^2)^2 + 4 \sigma_R^2 \sigma_I^2 \rho^2 } \, A^2  }{2 E_{0}^2 (1-\rho^2) \sigma_R^2 \sigma_I^2 g }  \right).
\label{hjvbq}
\end{align}
Now we introduce an alternative parametrization of the constants that depend on $\sigma_R$, $\sigma_I$, and $\rho$. Let us define the positive variables
\begin{align}
  X &= \frac{A}{E_0 \sigma_R \sigma_I \sqrt{1-\rho^2}} \cos(\omega), \nonumber \\
  Y &= \frac{A}{E_0 \sigma_R \sigma_I \sqrt{1-\rho^2}} \sin(\omega), \nonumber
\end{align}  
where the polar angle $\omega \in [0,\pi/2]$ is given by Eq. (\ref{kkop}). One can check that $X$ and $Y$ fulfill
\begin{equation}
X^2 + Y^2 = \frac{A^2}{ E_{0}^2 (1-\rho^2) \sigma_R^2 \sigma_I^2 } \nonumber
\end{equation}
and
\begin{equation}
X Y = \frac{\sqrt{(\sigma_R^2 - \sigma_I^2)^2 + 4 \sigma_R^2 \sigma_I^2 \rho^2 } \, A^2  }{2 E_{0}^2 (1-\rho^2) \sigma_R^2 \sigma_I^2 }, \nonumber 
\end{equation}  
which allows us to rewrite Eq. (\ref{hjvbq}) as follows
\begin{align}
&\mathcal{P}_{\beta}(A) = \frac{A}{E_{0}^2 \sigma_R \sigma_I \sqrt{1 - \rho^2}} \frac{n^n}{2^{n-1} (n-1)!}
\int_{0}^{\infty} dg \, g^{n-2} e^{- \frac{n g}{2}} \nonumber \\
& \times \exp{\left[- \frac{(X^2 + Y^2)}{2 g }  \right]} I_0 \left( \frac{X Y}{g} \right).
\label{hjvbq1}
\end{align}

Our aim is to find a convenient way to calculate the integral over $g$ in Eq. (\ref{hjvbq1}). This integral can be
seen as the ($n-2$)-th integer moment of the variable $g$, whose unormalized distribution is a product of an exponential and
a modified Bessel function of the first kind. The idea is to express higher-order moments ($n>1$) in terms of derivatives
of the lowest-order moment ($n=1$). This is a standard technique in statistical physics, which is implemented
here by defining the following function
\begin{equation}
H_n (u) = \int_{0}^{\infty} dg \, g^{n-2}  \exp{\left(- \frac{u g}{2}  - \frac{(X^2 + Y^2)}{2 g }  \right)} I_0 \left( \frac{X Y}{g} \right)
\end{equation}  
of the variable $u \in \mathbb{R}^+$. The integral appearing in Eq. (\ref{hjvbq1}) is recovered by calculating the function $H_n (u)$ at $u=n$. Thus, by
defining $H(u) : = H_1(u)$ and noticing that
\begin{equation}
\frac{\partial^{n-1} H(u)}{\partial u^{n-1} } = \left(- \frac{1}{2} \right)^{n-1} H_n (u) \,\,\, (n > 1),
\end{equation}  
we rewrite Eq. (\ref{hjvbq1}) as
\begin{equation}
  \mathcal{P}_{\beta}(A) = \frac{A}{E_{0}^2 \sigma_R \sigma_I \sqrt{1 - \rho^2}} \frac{(-1)^{n-1} n^n}{ (n-1)!}
  \frac{\partial^{n-1} H(u)}{\partial u^{n-1} } \Bigg{|}_{u=n}.
\label{hjvbq13}
\end{equation}
The above equation becomes a powerful identity to compute $\mathcal{P}_{\beta}(A)$ only if we are able to
solve the integral
\begin{equation}
H(u) = \int_{0}^{\infty} \frac{dg}{g}  \exp{\left(- \frac{u g}{2}  - \frac{(X^2 + Y^2)}{2 g }  \right)} I_0 \left( \frac{X Y}{g} \right).
\end{equation}  
Fortunately, it is possible to calculate the above integral using \cite{gradshteyn1}. The result is given by Eq. (\ref{hspopo}), which demonstrates Eq. (\ref{der}).

\bibliography{biblio.bib}

\end{document}